# An emergent quasi-2D metallic state derived from the Mott insulator framework

P.-C. Chiang,[1] S. C. Lin,[1] C.-Y. Chiang,[2] C.-S. Ku,[2] S. W. Huang,[3,4*] J. M. Lee,[4] Y.-D. Chuang,[5] H. J. Lin,[2] Y. F. Liao,[2] C.-M. Cheng,[2,6] S. C. Haw,[2] J. M. Chen,[2] Y.-H. Chu,[7,8*] T. H. Do,[9] C. W. Luo,[9,10] J.-Y. Juang,[9,10] K. H. Wu,[9] Y.-W. Chang,[10] J.-C. Yang,[10] and J.-Y. Lin[1,9*]

[1]*Institute of Physics, National Yang Ming Chiao Tung University, Hsinchu 30010, Taiwan*

[2]*National Synchrotron Radiation Research Center, Hsinchu 30076, Taiwan*

[3]*Swiss Light Source, Paul Scherrer Institut, 5232 Villigen PSI, Switzerland*

[4]*MAX IV Laboratory, Lund University, P. O. Box 118, 221 00 Lund, Sweden*

[5]*Advanced Light Source, Lawrence Berkeley National Laboratory, Berkeley, CA 94720*

[6]*Department of Physics, National Sun Yat-Sen University, Kaohsiung 80424, Taiwan*

[7]*Department of Materials Science and Engineering, National Yang Ming Chiao Tung University, Hsinchu 30010, Taiwan*

[8]*Center for Emergent Functional Matter Science, National Yang Ming Chiao Tung University, Hsinchu 30010, Taiwan*

[9]*Department of Electrophysics, National Yang Ming Chiao Tung University, Hsinchu 30010, Taiwan*

[10]*Department of Physics, National Cheng Kung University, Tainan 701, Taiwan*

**Abstract**

**Recent quasi-2D systems with judicious exploitation of the atomic monolayer or few-layer architecture exhibit unprecedented physical properties that challenge**

**the conventional wisdom on the condensed matter physics. Here we show that the infinite layer SrCuO$_2$ (SCO), a topical cuprate Mott insulator in the bulk form, can manifest an unexpected metallic state in the quasi-2D limit when SCO is grown on TiO$_2$-terminated SrTiO$_3$ (STO) substrates. The sheet resistance does not conform to Landau's Fermi liquid paradigm. Hard x-ray core-level photoemission spectra demonstrate a definitive Fermi level that resembles the hole doped metal. Soft x-ray absorption spectroscopy also reveals features analogous to those of a hole doped Mott insulator. Based on these results, we conclude that the hole doping does not occur at the interfaces between SCO and STO; instead, it comes from the transient layers between the chain-type and the planar-type structures within the SCO slab. The present work reveals a novel metallic state in the infinite layer SCO and invites further examination to elucidate the spatial extent of this state.**

**I. Introduction**

The discovery and study of emergent low-dimensional electronic systems have been among the recent focuses in condensed matter research activities. Seminal examples like graphene and two-dimensional (2D) materials, in particular with their twisted bilayer forms, can host unexpected Mott physics and even the superconductivity. [1-7] Parallel efforts on the interface engineering of complex oxides have generated wonders in several cases. [8-13] Noticeably, a 2D superconductivity in the context of two-dimensional electron gas (2DEG) at the interface between KTaO$_3$ substrate and EuO or LaAlO$_3$ overlayers was recently reported. [14] For exploring unconventional superconductivity, monolayer FeSe and CuO$_2$ have attracted much attention. [15,16] Monolayer FeSe on SrTiO$_3$ shows an enhanced superconducting transition temperature ($T_c$) of ~ 60 K from ~ 8 K in the bulk, whereas the monolayer $CuO_2$ plane likely hosts an *s*-wave rather than a *d*-wave order parameter as in the bulk high-$T_c$ superconductors. These results all suggest that innovative approaches to reach the limit of monolayer or a few atomic layers can generate a wide variety of quasi-2D systems displaying previously unachievable electronic properties.

The compound used in this study is $SrCuO_2$ (SCO). Infinite-layer (IL) tetragonal (planar-type) SCO consisting of 2D $CuO_2$ planes alternating with Sr layers is the simplest structure among high-$T_c$ cuprates. The recent discovery of superconductivity in infinite layer nickelate thin films show the rich phenomena that may be induced in such a structure. [17] However, bulk IL SCO can only be synthesized under high pressure. [18] Recently, extensive experimental efforts have sought to elucidate the structure features of tetragonal SCO in thin film fabrications. [19-22] First principles calculations indicated a structural transformation from chain-type to planar-type with increasing SCO film thickness (> 5 unit cells (u.c.)). [23] (Both structures are shown in the schematic picture in Fig. 1(a).) Apart from the studies of bulk properties, tetragonal SCO also serves as an essential building block for super-lattice engineering to compose "artificial" high-$T_c$ compounds. [24-29] In this paper, we present the discovery of an emergent metallic state in ultrathin SCO films grown on $TiO_2$-terminated $SrTiO_3$ substrates.

## II. Methods

Ultrathin $SrCuO_2$ films were fabricated on $TiO_2$-terminated (001) $SrTiO_3$ (STO) substrates (SCO/STO) using the pulsed-laser deposition (PLD) method. The substrate was first treated with the HF-$NH_4$F buffer solution to produce a uniform $TiO_2$ termination. The layer-by-layer growth of SCO was ensured by monitoring the clear intensity oscillations in the in-situ reflection high-energy electron diffraction (RHEED), see Figs. 1(c) and 1(d). We have fabricated nearly 100 SCO/STO ultrathin films with varied growth parameters and thicknesses. In this paper, we focus mainly on samples deposited at about 500ºC under 210 mTorr oxygen pressure. To obtain the films that display a metallic behavior, it is crucial to control all growth parameters to within 5% variation. Albeit with such an elaborated control of growth condition, only ~ 10% of films among these focused samples show the decreasing sheet resistance $R_\square(T)$ with decreasing temperature $T$. Importantly, *all* of the focused films exhibits a nearly metallic behavior with at least three orders of magnitude larger sheet conductance than that of IL SCO thin film, as seen in Fig. S3 in the Supplementary Materials (SM).

Polarization-dependent x-ray absorption spectroscopy (XAS) at both Cu *L*-edges and O *K*-edge was carried out at beamlines 11A and 20A at National Synchrotron Radiation Research Center (NSRRC) in Taiwan with 0.12 eV and 0.2 eV energy resolutions, respectively. The spectra were recorded in the total electron yield mode (TEY, sample-to-ground drain current) for all SCO/STO ultrathin films. For IL SCO films that are insulating (thickness ~ 300 nm), XAS spectra were recorded in the total fluorescence yield mode (TFY) using a microchannel plate detector. The thin films were mounted with Cu-O bond direction in the horizontal scattering plane. The *c*-axis of film was parallel to the Poynting vector of x-rays in the normal incidence geometry. Rotating the *c*-axis in the horizontal plane varied the projection of linear horizontal polarization (π-polarization) of electric field from **E**//*ab* (normal incidence) to **E**//*c* (30º grazing incidence; x-ray polarization mostly along the *c*-axis). The TFY mode of XAS is bulk sensitive while the TEY mode is very surface sensitive. The probe depth of TEY mode of XAS at O-*K* edge was carefully determined to be ~ 8

u.c. in SCO (see Fig. S6 for the thickness dependent XAS spectra; the STO substrate signal is visible in the 5 u.c. but not the 8 u.c. spectra).

Laue x-ray nano-diffraction measurements were performed at beamline 21A at Taiwan Photon Source (TPS) of NSRRC. Using the state-of-the-art x-ray focusing components, TPS 21A is capable of providing a tightly focused x-ray beam with a spatial resolution of 70 nm and the energy range from 5 to 30 keV. Unlike typical x-ray diffraction (XRD) measurement where a monochromatized x-ray is used for scattering and the sample and detector angles are scanned to record the diffraction peaks, we fixed the sample and detector angles and varies the incident x-ray energy from 8 to 20 keV. This is because the $c$-axis lattice constants of STO (0.3905 nm) and the chain-type SCO (0.3895 nm) are nearly identical to each other such that they cannot be resolved with the resolution of conventional XRD setup.

The hard x-ray core-level photoemission measurements (HAXPES) were carried out at the Taiwan beamline BL12XU at Spring8, Japan. The photon beam is linearly polarized with the electric field vector in the horizontal plane. Photoelectrons were detected by a MB Scientific A-a HE analyzer mounted parallel to the electrical field of photon with an overall resolution of 0.28 eV. [29]

The sheet resistance $R_\square$ of the SCO ultrathin films and the room temperature Hall measurements were implemented using the van der Pauw method. To prevent the degradation of these ultrathin films, it is crucial to limit the measurement current. Typically, a current of 50 pA is deployed.

To supplement the spectroscopic measurements, we also carried out the scanning transmission electron microscopy (STEM) on selected SCO/STO films. The STEM images were taken using JEOL ARM200F equipped with a spherical aberration (Cs) corrector at 200 kV accelerating voltage. The semi convergent angle was 25 mrad, which formed a < 1 Å electron probe. The semi collection angle of high-angle annular dark-field (HAADF) detector was 68 to 280 mrad.

### III. Results and Discussions

Some of the SCO/STO ultrathin films with growth parameters mentioned above exhibit a metallic behavior. For example, the $R_\square(T)$ of one sample is shown in Fig. 2(a). For this sample and others alike, $R_\square(T)$ decreases with decreasing $T$ in the temperature range from 300 K to 2 K, showing the characteristics of a metal. The result is intriguing because the bulk IL SCO is considered a Mott insulator. The metallic state in SCO/STO ultrathin films can be reliably reproduced, see Fig. S1 in Supplemental Materials (SM). A fundamental question about this emergent metallic state is the type of carriers participating in the transport properties. To answer this question, we performed the Hall measurements on these samples. The positive slope of Hall resistance $R_{\text{Hall}}(H)$ in Fig. 2(b) clearly indicates the hole nature of these carriers. Fig. 2(c) shows the valence band HAXPES of a 15 u.c. SCO/STO measured at 300 K. The Fermi level (zero binding energy) was determined by a gold film next to the sample. The finite spectral weight at Fermi level reflects a metallic state in this SCO/STO sample. Besides, the incident photon energy of HAPES was set to ~ 6.5 keV to ensure a large probe depth to reveal the metallic state lying deep inside the SCO slab, which is the key to the scenario proposed in the following sections. [30] As a short summary, the experimental results in Fig. 2 represent the core discovery of an unconventional metallic state in SCO/STO.

The nature of these hole carriers can be further explored by using soft x-ray absorption spectroscopy (XAS), which directly probes the unoccupied electronic states. Fig. 3(a) shows the Cu $L_{\text{III}}$ edge XAS spectra of the reference bulk IL SCO (100 nm thick). The strong anisotropy seen in the spectra is related to the orbital configuration of Cu in the $CuO_2$ planes, and the enhanced spectral intensity with photon polarization in the film surface (**E**//*ab*) is consistent with the planar lattice structure of IL SCO (see Fig. 1(a)). Similar anisotropy can also be seen in the metallic SCO/STO (Fig. 3(b)); however, close inspections do reveal some crucial differences between the XAS spectra in Figs. 3(a) and 3(b). Specifically, in contrast to the symmetric peak profile at Cu $L_{III}$ edge (~ 931.5 eV) for bulk IL SCO (a Mott insulator), the corresponding peak in Fig. 3(b) for metallic SCO gains additional spectral weight on the high-energy side around 933 eV, which is often attributed to the ligand hole states in doped Mott insulators. [31,32]

Both upper Hubbard band (UHB) and Zhang-Rice (ZR) singlets are the essential features of Mott physics. Due to the multiplicity effects on Cu *L*-edge, both UHB and ZR singlets are more clearly revealed in O *K*-edge XAS spectra through Cu 3*d* and O 2*p* orbital hybridization. [31,32] We note that unlike Cu *L*-edge XAS, O *K*-edge XAS was rarely performed on SCO-related compounds. In Fig. 3 (c), we present the O *K*-edge XAS spectra of both IL SCO and metallic 15 u.c. SCO/STO. Since the IL SCO is an undoped Mott insulator, the dominant peak in the pre-edge region can be assigned to the UHB. For metallic SCO/STO, a substantial spectral weight appears at even lower energies, which is related to the ligand hole states seen in Cu *L*-edge XAS spectrum (black line, Fig. 3(b)). The results in Fig. 3(c) show that the hole carriers in metallic SCO/STO have a dominant O 2*p* character just like the hole-doped cuprates superconductors. We used three Gaussian components to fit the O *K*-edge XAS spectrum of metallic SCO/STO to determine the energies of these features. The two dominant peaks from the fitting are denoted by the dashed lines in Fig. 3(c). The third component, reminiscent of the tail of high energy feature, is not shown for clarity. From the fitting, we see the higher energy peak is at nearly the same energy as that of the UHB whereas the lower energy peak associated with doped holes is ~ 1 eV lower than the UHB, matching the energy difference between ZR singlets and UHB in other cuprates. [32] It is thus reasonable to assign the lower energy peak to the ZR band in metallic SCO and the holes to the ZR singlets.

It is also prudent to verify that the doped holes exhibit the dominant O $2p_{x,y}$ character as anticipated for ZR singlets. This is confirmed by the strong anisotropy in the spectrum in Fig. 3(d) where the enhanced spectral intensity is seen with **E**//*ab* (black curve). Although the spectral weight is severely suppressed with **E**//*c* (red curve), remnant weight is still visible in the spectrum. This *c*-axis component is also consistent with the observed (but small) ligand-hole spectral weight about 933 eV in the Cu *L*-edge XAS spectrum with **E**//*c* (red curve, Fig. 3(b)). The **E**//*c* component in both O *K*-edge and Cu *L*-edge XAS is due to apical oxygen [21,32], a feature absent in IL SCO but *crucial* to our model for explaining this emergent metallic state.

We also utilized the Laue nano-diffraction to reveal more structure details of a 10 u.c. SCO/STO. Firstly, we applied white-beam x-rays to cover 5 - 30 keV energy range for a Laue diffraction pattern and indexed the crystal planes for each diffraction feature as shown in Fig. 4 (a). The (103) lattice plane of planar-type SCO can be clearly observed in the Laue patterns because of the very different *a*/*c* ratios between the planar-type SCO and STO substrate. However, it is difficult to distinguish the x-ray diffraction pattern of chain-type SCO from the STO substrate because they have nearly identical *a*/*c* ratios. To overcome this constraint, we performed an energy scan with a fixed detector angle ($2\theta$ = 95.7°) to cover both chain-type SCO and STO around (0 0 4) (8 to 9 keV) and (0 0 6) (12 to 13 keV). The results are shown in Fig. 4 (b). The *x*-axis is converted from photon energy to reciprocal lattice unit (r.l.u.) according to the Bragg's law. An additional shoulder-like feature clearly emerges at higher energy side, corresponding to a slightly smaller *c*-axis lattice constant (0.3895 nm for chain-type SCO versus 0.3905 nm for STO). This x-ray mono-beam energy-scan spectrum provides the direct evidence for the elusive chain-type SCO structure in a plain SCO thin film for the first time, and confirms the existence of residual chain-type layers within a supposed planar-type SCO/STO with SCO thickness greater than 5 u.c.

We also performed STEM measurement on a SCO/STO. In the STEM image (Fig. S7(a)), the positions of Cu and O atoms are clearly visible. To visually highlight the closeness of *c*-axis lattice constants between chain-type SCO and STO substrate, we use the same black box as the marquee to denote their unit cells. At the top most layer, one sees that the *c*-axis lattice constant is greatly reduced such that the marquee becomes a rectangle. The *c*-axis lattice constant can be obtained from fitting this STEM image (Fig. S7(b)). The fitting shows that the first SCO layer next to STO has $c = 0.388$ nm, close to the 0.3895 nm determined from Laue nano-diffraction (Fig. 4(b)) and is slightly smaller than $c = 0.3905$ nm of the STO substrate. It drops to $c \approx 0.350$ nm beyond first three layers, close to 0.343 nm for the bulk planar-type SCO. These results imply that the first few layers of SCO at the interface will have a chain-type structure, corroborating with the findings from Laue nano-diffraction. They also generally support the model in Fig. 4 (c) as we shall see.

A comprehensive understanding of this newly found metallic state is currently unavailable, but two scenarios can be excluded. (i) Since STO is non-polar and the planar-type SCO is polar, one might speculate that this metallic state could resemble the well known 2DEG at the interface of LaAlO$_3$/SrTiO$_3$. [6] Nevertheless, the hole nature of carriers in the present system is not reconciled with the mechanism of polar catastrophe or the electronic reconstruction at the interface of LaAlO$_3$/SrTiO$_3$.

Moreover, STEM and Laue nano-diffraction all suggest the presence of chain-type SCO in between the STO substrate and the planar-type SCO once the SCO thickness exceeds 5 u.c. (Fig. 4 (b) and S7). In that regard, the polar planar-type SCO will not be in proximity to the STO substrate. (ii) Intuitively, one might suspect that the metallic state should reside at the interfaces between SCO and STO. However, $R_\square$ (300 K) of a 2 u.c. SCO/STO is so large that it is beyond our measurement capability (Fig. S4). The $R_\square$(300 K) of 5 u.c. SCO/STO is also larger than that of the 8 ~ 15 u.c. SCO/STO films. Corresponding changes in the electronic structure can also be seen in the O $K$-edge XAS spectra. In Fig. S6, the ZR singlet signature is absent in the 2 u.c. SCO/STO. It becomes visible once the SCO thickness reaches 5 u.c. and remains visible up to 15 u.c. (Fig. 3(c)). These results indicate that the interfacial region between chain-type SCO and STO substrate is not metallic.

Based on the transport, structure, and soft x-ray spectroscopic results, we propose a model for this emergent metallic state, which is illustrated in Fig. 4(c). This proposed model might not be an exact one, but it can serve as an enlightening first step to explore this emergent state. For SCO ultrathin films grown on STO, where the epitaxial strain plays an important role, it was predicted that a structural transformation in SCO from chain-type to planar-type would take place when the SCO thickness exceeds 5 u.c.. [23] This is experimentally confirmed in this study and is consistent with literature. [26] However, we suggest a more detailed context based on the Laue nano-diffraction and STEM results. Near the STO interface, the initial growth of SCO will have a chain-type structure. With increasing SCO thickness, there exist residual layers of chain-type SCO even when the planar-type structure has developed. Within the chain-type SCO, SrO layers alternate with CuO layers (lower part of Fig. 4(c)). But in the planar-type, Sr layers exist with $CuO_2$ layers (upper part of Fig. 4(c)). It is plausible that, as the key to this model, there exist SrO$_x$ layer(s) *in the transient regime* (highlighted by the block at the middle of Fig. 4(c)) in which excess oxygen is located on top of Cu in the neighboring CuO$_2$ plane(s). In this context, the features of Cu *L*-edge and O *K*-edge XAS in Figs. 3(a) - 3(d) can be explained: the excess oxygen in SrO$_x$ layers effectively dopes the holes into the nearby CuO$_2$ planes. The observed conductivity hence arises from the few doped CuO$_2$ planes similar to that in cuprate superconductors. The quasi-2D nature of the hole states observed in Fig. 3(d) is thus a natural aspect of this model. This model is also consistent with that the thick (such as 100nm films) film shows the insulating behavior in the sheet resistance $R_\square(T)$ measurements, since the applied current of present top electrode configuration only probes the upper insulating planar-type SCO layers while the metallic layer is buried deeply below the surface.

A previous study of Ca doping in YBa$_2$Cu$_3$O$_y$ (YBCO) indicated that the doped holes due to Ca ions are distributed only in the neighboring CuO$_2$ planes. [33] As the location of SrO$_x$ layer in our model is similar to that of Ca ions in Y$_{1-x}$C$_x$aBa$_2$Cu$_3$O$_y$, we suppose that the doped holes due to apical O$_x$ are confined mainly to the next CuO$_2$ plane. It is worth comparing the current doping mechanism with that of "long-*c*" phase Sr$_{1-x}$La$_x$CuO$_{2+y}$ (SLCO). [34-36] Structure-wise, the *c*-axis of "long-*c*" phase SLCO ranges from 0.360 nm to 0.365 nm, which is apparently shorter than the *c*-axis of residual chain-type SCO layers (see Fig. 4(b)). While the resultant doping in "long-*c*" phase SLCO is a combined effect of both La$^{3+}$ (electron doping) and apical oxygen (hole doping), the hole doping mechanism depicted in Fig. 4(c) comes from apical oxygen in the transient layer(s). The latter mechanism is simpler and involves only hole-type carriers, thus it could potentially serve as a controllable way for the manipulation of present metallic state.

In addition to the identification of hole-doped CuO$_2$ planes in SCO/STO ultrathin films, we further discuss some of the pivotal elements crucial for the underlying physics. $R_\square(T)$ in Fig. 2(a) explicitly reveals two essential features of this metallic state: it is a clean metal according to the large residual resistance ratio (RRR). The absence of a flat regime at the lowest temperatures in $R_\square(T)$ rules out a dominant contribution from an electron-phonon interaction to the temperature dependence of $R_\square(T)$. Moreover, there is no signature of $T^2$ dependence due to electron-electron interaction at low temperatures as expected in a low-dimensional Fermi liquid.

$R_\square(T)$ in Fig. 2(a) is much lower than the threshold $R_\square \approx h/e^2 = 25.8$ k$\Omega$ for 2D metals. Regarding the number of hole carriers in this metallic state, the direct conversion of Hall resistance $R_{\text{Hall}}$ from Fig. 2(b) leads to $1.5 \times 10^{14}$ hole/cm$^2$ or a reasonable doping level of $p = 0.23$/unit cell assuming a single conducting CuO$_2$ plane. Previous attempts to use Hall coefficients to estimate the carrier density in cuprate superconductors have shown that despite having the correct sign for the carrier type, they quantitatively deviate from the actual hole doping level. [41,42] At this moment, a reliable theory for the relation between the Hall coefficient and the carrier number in cuprates is still unavailable. Moreover, without knowing the exact number of conducting CuO$_2$ layers in the system, one can only have qualitative discussions on the doping level (the number of conducting CuO$_2$ layers is discussed in SM). It is intriguing to note that the temperature dependence of $R_\square(T)$ in Fig. 2(a) differs from that of over doped cuprates. [32,37,38] Hole-doping methods in the literature for cuprates have never been able to achieve a doping level much larger than $p = 0.3$. [34,39] Doping mechanism based on intrinsic strain-induced structural transition in the present scenario likely offers an alternative route to the unprecedented hole-doping regime. The deep doping regime can be the venue of unique properties unseen in previous over-doped cuprates. [40]

A pressing question is whether this metallic state can be tuned into a superconducting state or not. Based on the above scenario, a venue to create superconductivity would be to vary the doping level. A decrease in $p$ would make the system revert to the regime of a "conventional" high-$T_c$ cuprate. An increase in $p$ probably leads to an uncharted regime in which new superconductivity could revive. [37] Apparently there exist avenues to engineer the physical properties of this emergent state. For example, the control of $p$ can be realized with either judicious preparation of SCO/STO under various oxygen partial pressure or with an electric-field effect to adjust the Fermi level. Efforts to adapt both designs are currently in progress.

E-mail: shih.huang@psi.ch ; ago@nycu.edu.tw; yhchu@mx.nthu.edu.tw

**Acknowledgements**

We thank the technical help from H. Lin on STEM experiments. This work was supported by Taiwan MOST under grant Nos. 108-2112-M-009-009, MOST 110-2634-F-009-026, and107-2119-M-009-022, and by Center for Emergent Functional Matter Science of National Yang Ming Chiao Tung University from The Featured Areas Research Center Program within the framework of the Higher Education Sprout Project by Taiwan Ministry of Education financially supported the work of P. C. C. and J. Y. L.. Y.D.C is supported by the Advanced Light Source, a U. S. DOE Office of Science User Facility under contract no. DE-AC02-05CH11231.

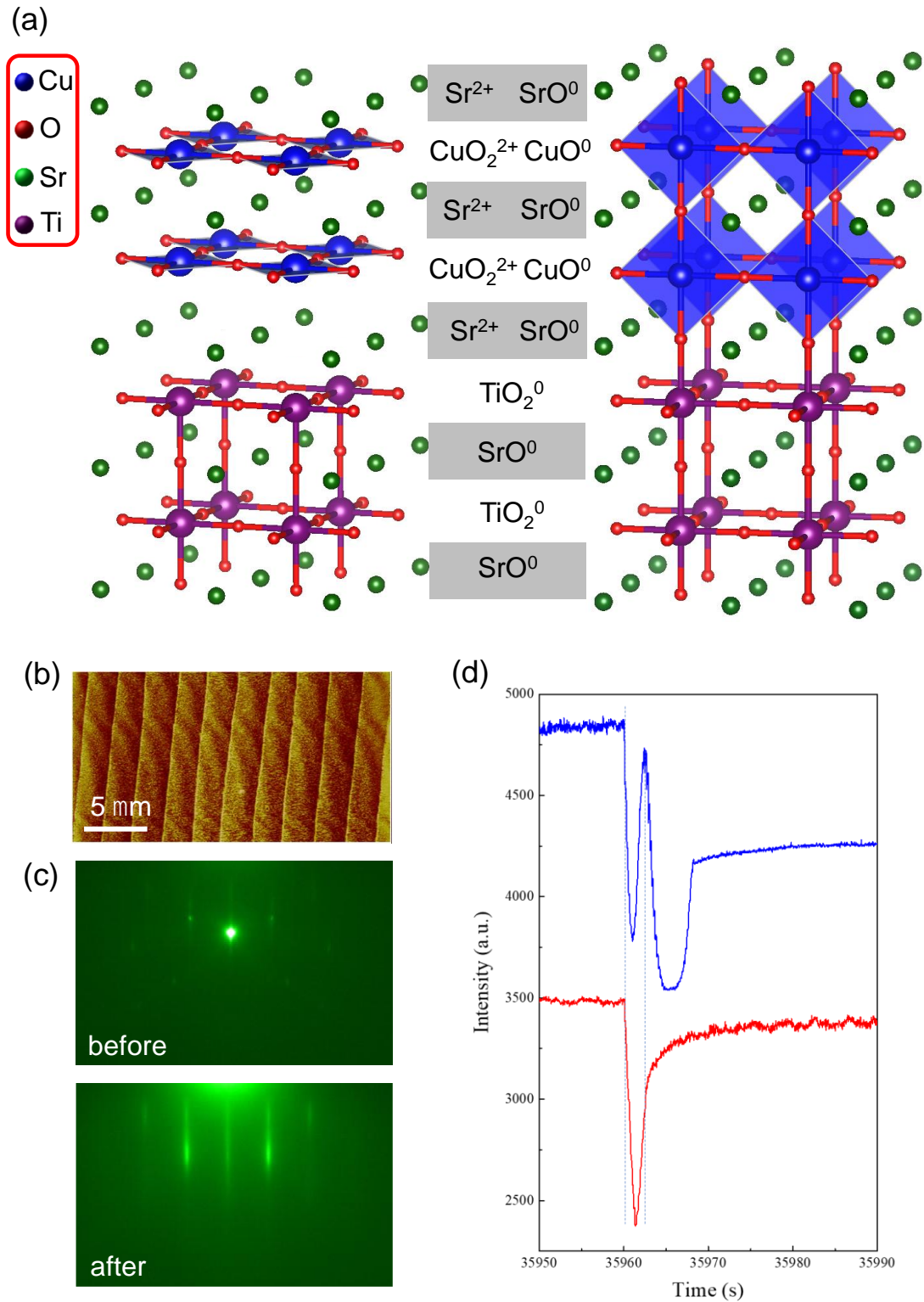

FIG. 1. (color online) (a) Schematic plots of the planar-type (left) and the chain-type (right) tetragonal SCO on $TiO_2$-terminated $SrTiO_3$ substrates. (b) Surface morphology of $TiO_2$-terminated $SrTiO_3$ substrates characterized by an atomic-force microscope.

The single unit-cell height steps can be clearly seen in the picture. (c) RHEED patterns before and after the growth of a monolayer SCO. (d) The intensity oscillations of the specularly diffracted spots in the RHEED pattern during the growth of a monolayer (red, bottom curve) and a 2 u.c. (top, blue curve) SCO ultrathin films.

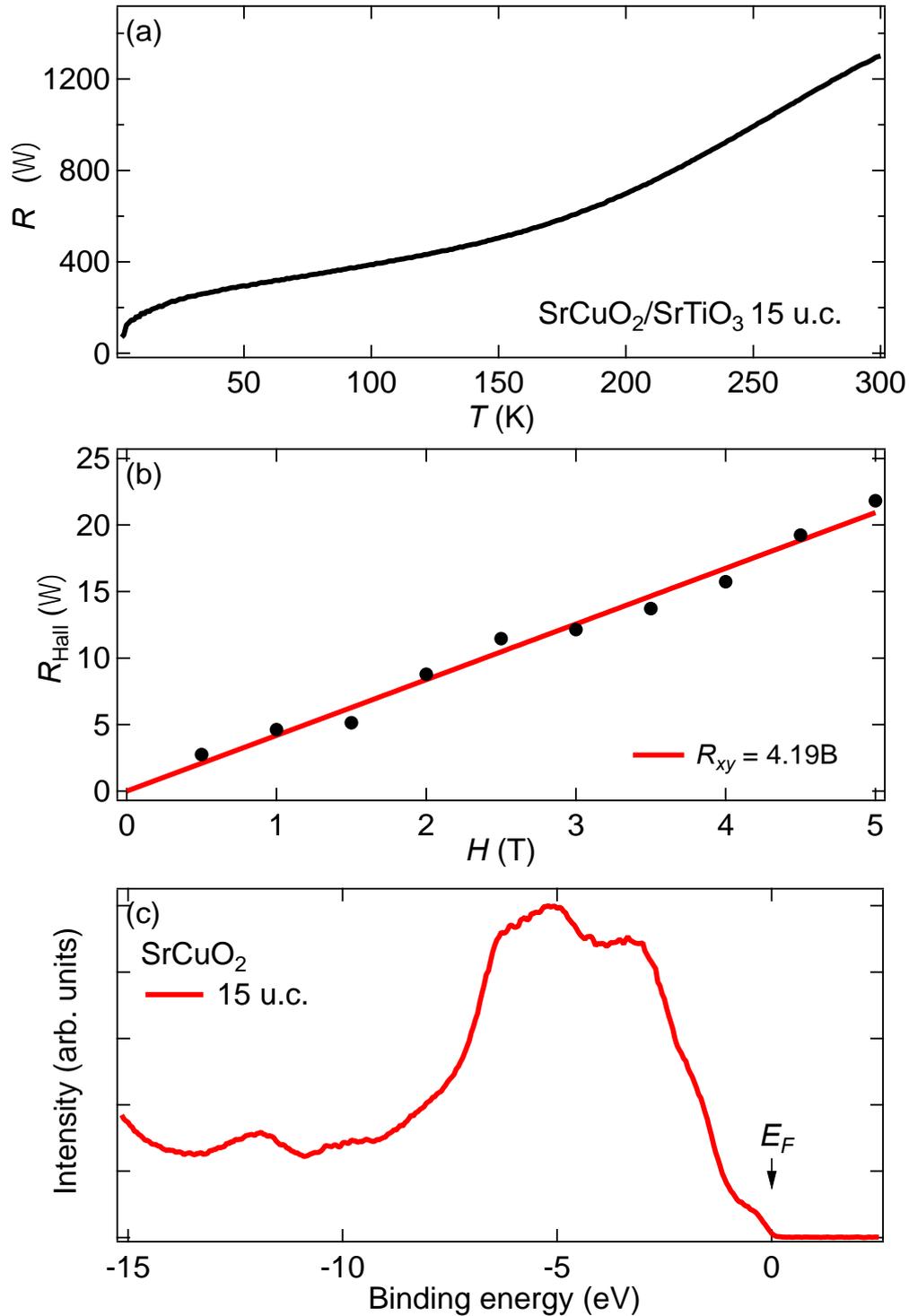

FIG. 2. (color online) (a) Sheet resistance $R_\square(T)$, (b) Hall resistance, and (c) the valence band HAXPES spectrum of a 15 u.c. SCO/STO (For discussions, see SM). The Hall and HAXPES measurements were performed at 300 K.

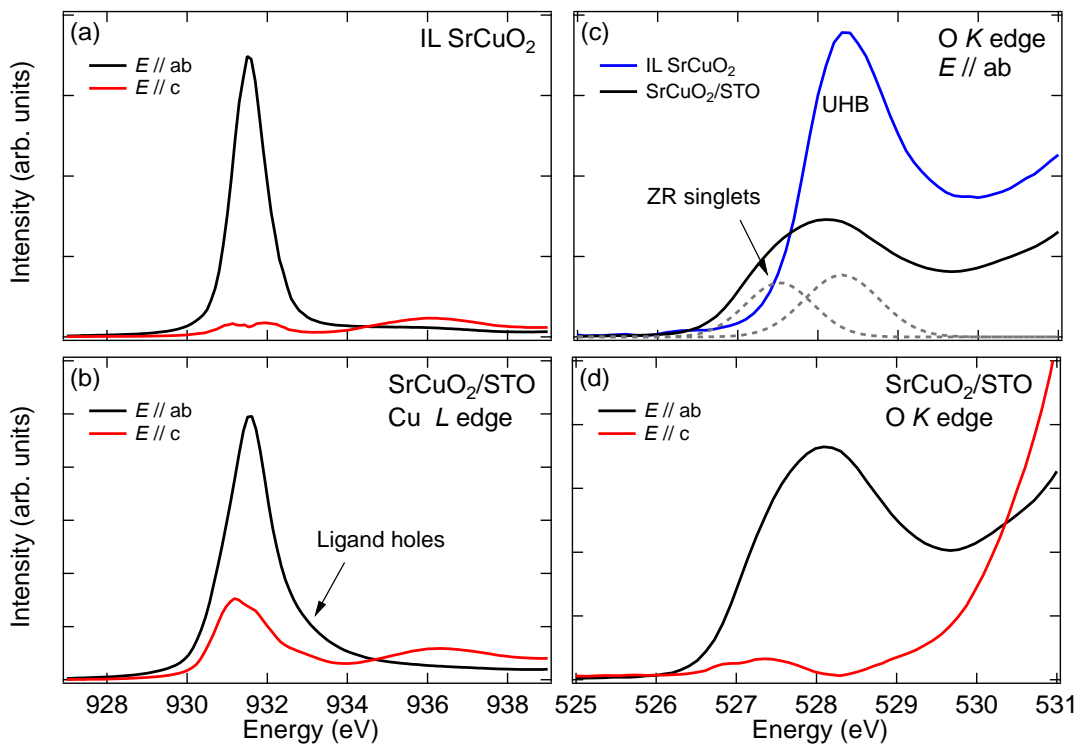

FIG. 3. (color online) Polarization dependence of Cu $L$-edge XAS spectra of (a) IL SCO and (b) the 15 u.c. SCO/STO. Black and red lines are spectra with **E**//$ab$ and **E**//$c$, respectively. (c) O $K$-edge XAS spectra of IL SCO and 15 u.c. SCO/STO; (d) Polarization dependence of O $K$-edge XAS spectra of 15 u.c. SCO/STO. The dashed lines in (c) denote the respective components for ZR singlets and UHB for SCO/STO.

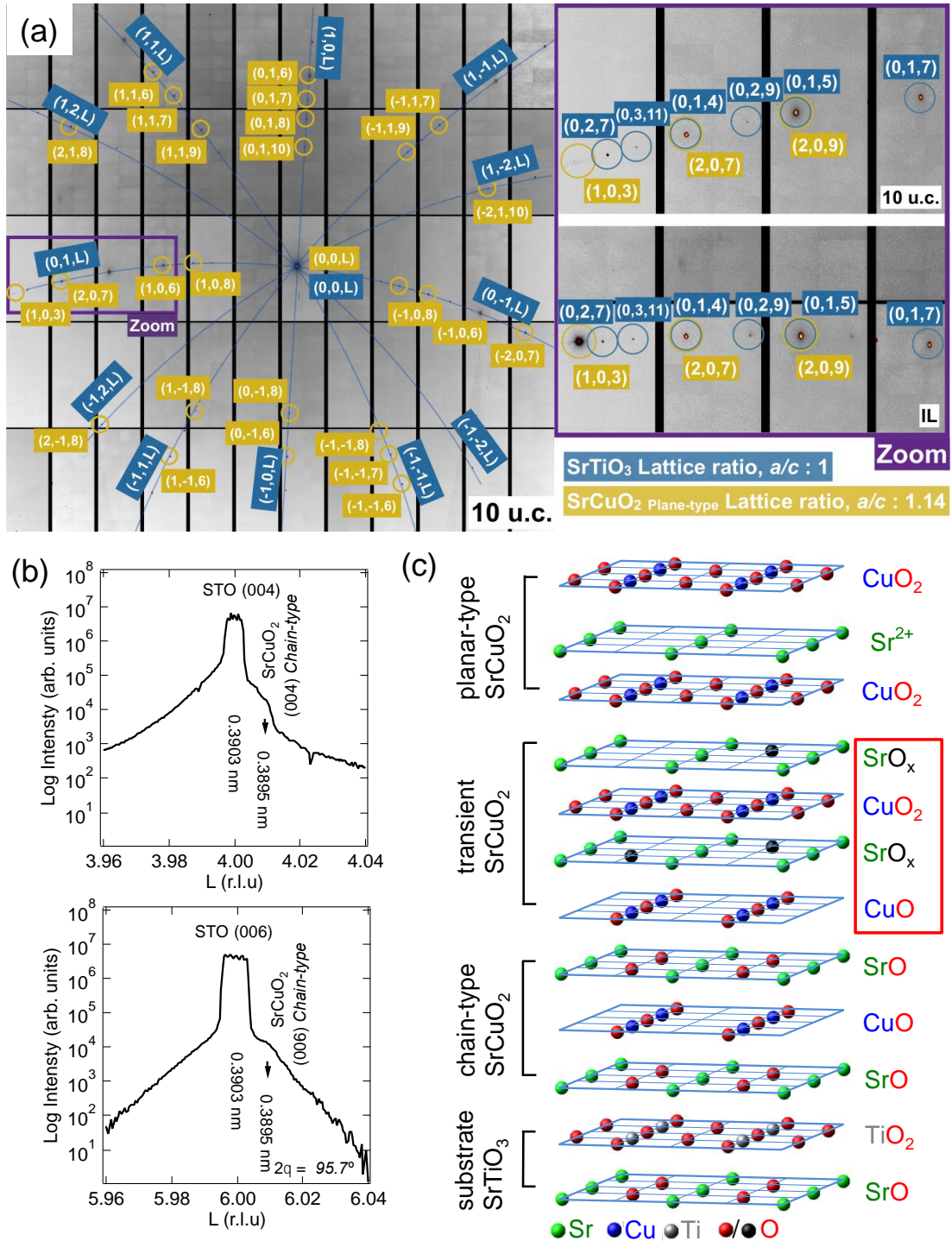

FIG. 4. (color online) (a) White-beam Laue diffraction pattern of a 10 u.c. SCO/STO sample. The diffraction peaks from STO substrate and the planar-type SCO are marked in steel blue and gold colors, respectively. The zoom in images (purple box) show the diffraction patterns of a 10 u.c. SCO/STO and an IL SCO. The (103) lattice plane of planar-type SCO can be clearly seen in the Laue patterns because of the different *a*/*c* ratio between planar-type SCO and STO substrate. (b) Energy scan of x-ray Laue diffraction. The reciprocal lattice unit (r.l.u.) plots for (004) and (006) lattice plane are produced by converting an energy scan (8 to 9 keV and 12 to13 keV with 1 eV energy step, respectively) of (00L) Laue diffraction peak with detector at 95.7º (two-theta angle). The existence of the chain-type SCO was revealed as a shoulder next to the strong STO peak. The *c*-axis lattice constant of chain-type SOC is determined to be 0.3895 nm. (c) Schematic of layered SCO/STO structure showing the transition region from chain-type to planar-type SCO. The metallic region and the doping mechanism are highlighted by the red box.

# Supplementary Materials for

# An emergent quasi-2D metallic state derived from the Mott insulator framework

P.-C. Chiang,[1] S. C. Lin,[1] C.-Y. Chiang,[2] C.-S. Ku,[2] S. W. Huang,[3,4*] J. M. Lee,[4] Y.-D. Chuang,[5] H. J. Lin,[2] Y. F. Liao,[2] C.-M. Cheng,[2,6] S. C. Haw,[2] J. M. Chen,[2] Y.-H. Chu,[7,8*] T. H. Do,[9] C. W. Luo,[9,10] J.-Y. Juang,[9,10] K. H. Wu,[9] Y.-W. Chang,[10] J.-C. Yang,[10] and J.-Y. Lin[1,9*]

[1]*Institute of Physics, National Yang Ming Chiao Tung University, Hsinchu 30010, Taiwan*

[2]*National Synchrotron Radiation Research Center, Hsinchu 30076, Taiwan*

[3]*Swiss Light Source, Paul Scherrer Institut, 5232 Villigen PSI, Switzerland*

[4]*MAX IV Laboratory, Lund University, P. O. Box 118, 221 00 Lund, Sweden*

[5]*Advanced Light Source, Lawrence Berkeley National Laboratory, Berkeley, CA 94720*

[6]*Department of Physics, National Sun Yat-Sen University, Kaohsiung 80424, Taiwan*

[7]*Department of Materials Science and Engineering, National Yang Ming Chiao Tung University, Hsinchu 30010, Taiwan*

[8]*Center for Emergent Functional Matter Science, National Yang Ming Chiao Tung University, Hsinchu 30010, Taiwan*

[9]*Department of Electrophysics, National Yang Ming Chiao Tung University, Hsinchu 30010, Taiwan*

[10]*Department of Physics, National Cheng Kung University, Tainan 701, Taiwan*

*E-mail: shih.huang@psi.ch ; ago@nycu.edu.tw; yhchu@mx.nthu.edu.tw

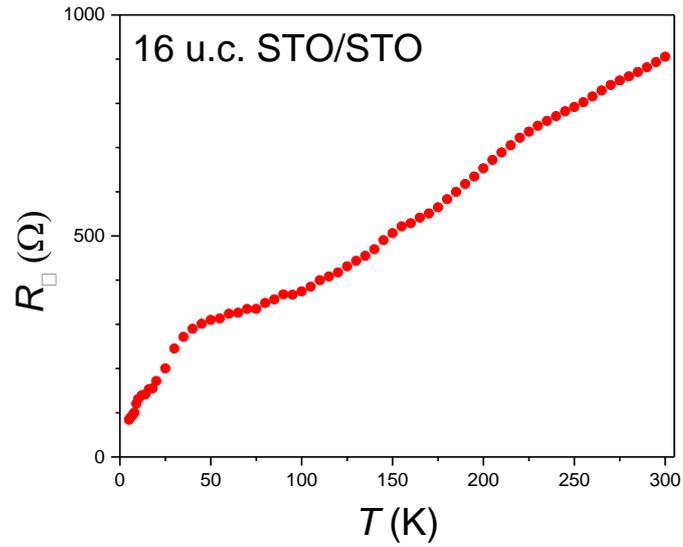

Fig. S1 **Reproducible metallic state.** Sheet resistance $R_\square(T)$ of another SrCuO$_2$/SrTiO$_3$ (SCO/STO, 16 u.c.) sample showing the metallic behavior. This sample was measured in a different setup (Quantum Design© electrical transport Option, ETO) compared to the ordinary 4-probe DC method used in Fig. 2 in the manuscript. Using two separate measurement setups to avoid systematic error, the metallic behavior was independently verified.

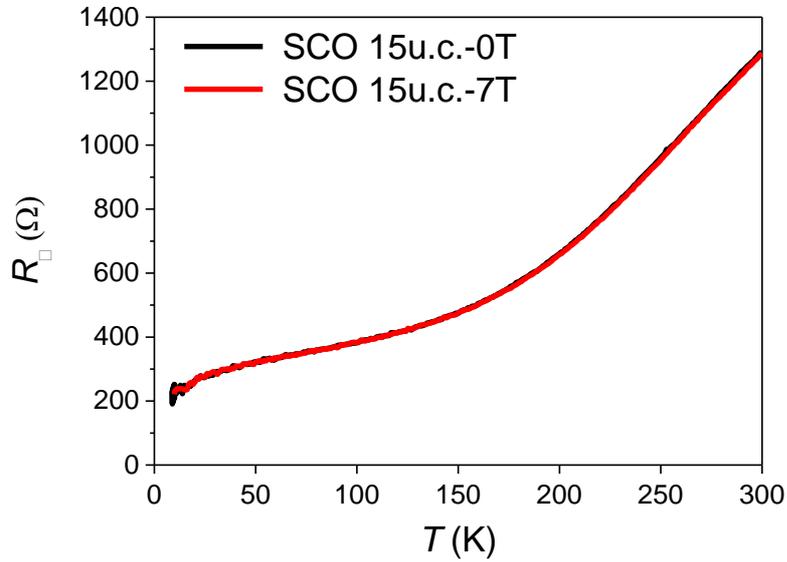

Fig. S2 **Absence of Magnetoresistance.** Sheet resistance $R_\square(T)$ of a 15 u.c. SCO/STO measured at zero field (black line) and 7 T field (red line). There is no appreciable magnetoresistance in this sample.

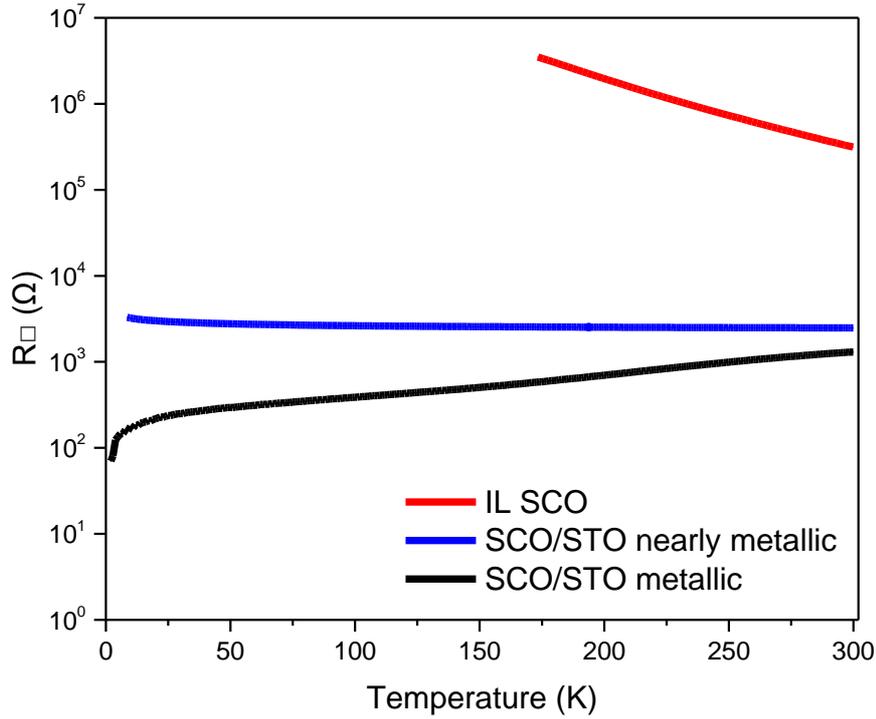

Fig. S3 **Comparison of the sheet resistance $R_\Box(T)$ of a metallic SCO/STO (from Fig. 2(a) in the paper), a nearly metallic SCO/STO, and a Mott insulating infinite-layer (IL) SCO.** Many fabricated SCO/STO ultrathin films show a nearly metallic behavior like the blue line from an 8 u.c. SCO/STO film. With fine tuning the growth parameters, we can improve the conductivity and roughly 10% of all samples will eventually show the metallic behavior like the black line here (Fig. 2(a) in the paper). For a Mott insulating IL SCO (100 nm, red line), the $R_\Box(T)$ value is too large to be measured accurately below 175K, hence the low temperature data points are not shown here. The $R_\Box(T)$ curves of metallic samples do not show a significant thickness dependence, see Fig. S4.

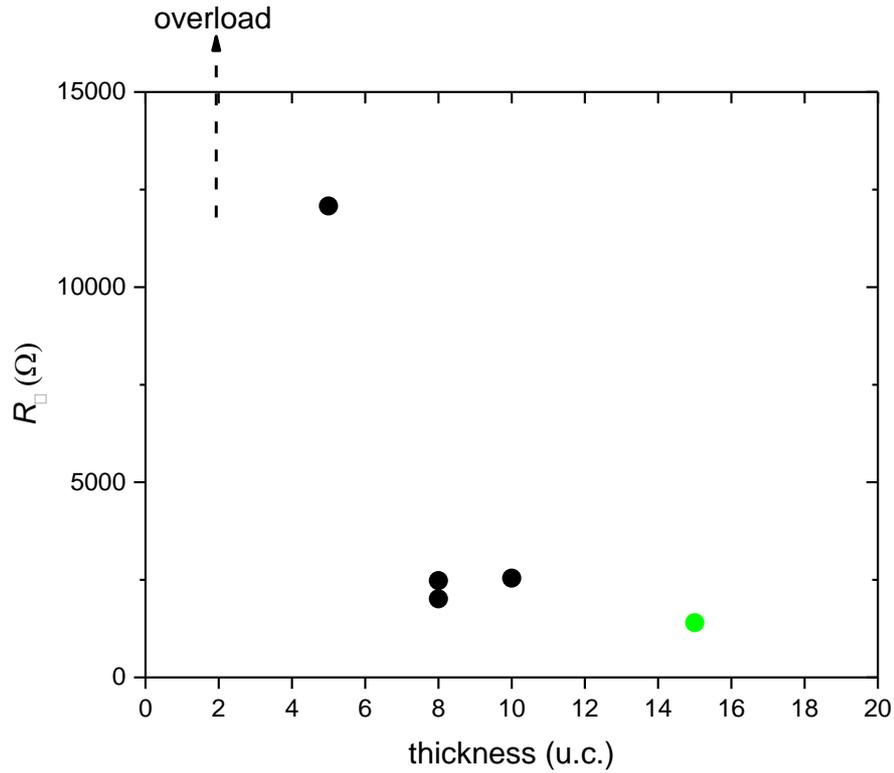

Fig. S4 **Thickness dependence of $R_\square$ at 300K.** The black solid circles are the $R_\square$ values of one batch of SCO/STO ultrathin films with thickness ranging from 2 u.c. to 10 u.c. $R_\square$ of 2 u.c. sample is too large to be measured accurately even at the room temperature. The green solid circle denotes the $R_\square$ value of the metallic SCO/STO (15 u.c.) shown in the manuscript.

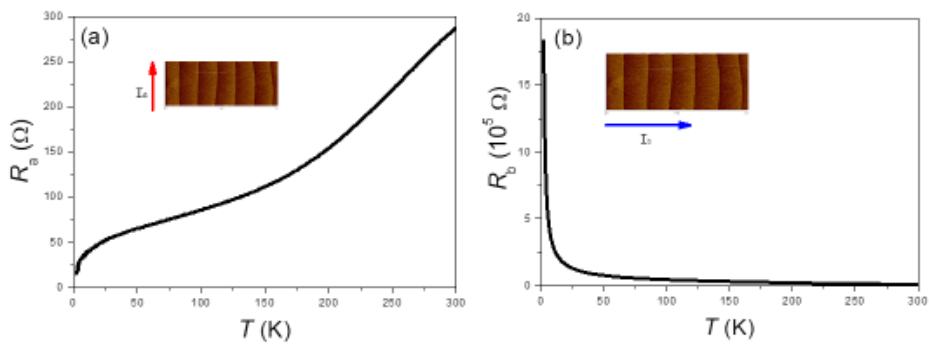

Fig. S5 **Anisotropic $R_\square(T)$ of the metallic SCO/STO.** Metallic SCO/STO samples similar to the one in Fig. 2(a) in the manuscript all show a large anisotropy with respect to the direction of applied current ($I_b$). We believe this anisotropy is due to steps on the substrates as revealed in the AFM images (inset). When the direction of applied current $I_b$ is nearly perpendicular to the steps (right panel), these samples

show an insulating behavior. On the other hand, when the direction of $I_b$ is nearly parallel to the steps (left panel), free from the influence of steps, the resistance $R_a$ shows a metallic behavior. We then assume $R_a$ in Fig. S2(a) is intrinsic and obtain $R_\square(T)$ in Fig. 2(a).

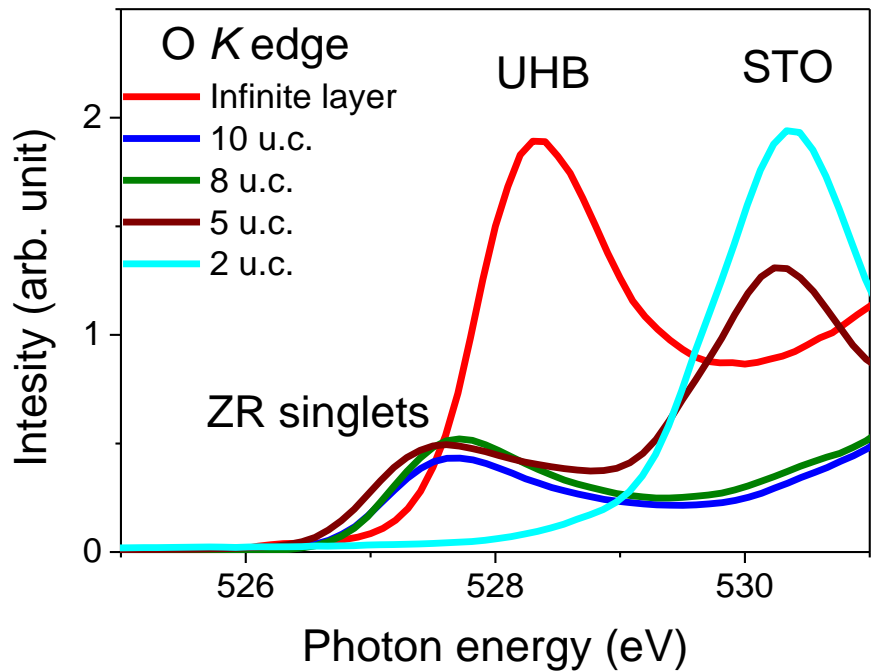

Fig. S6 **O $K$-edge XAS of SCO/STO with various SCO thickness.** The presence of Zhang-Rice (ZR) singlet in SCO with thickness larger than 5 u.c. generally coincidences with a significant electrical conductivity (see Figs. S1 and S3). The feature near 530.3 eV, clearly seen in 2 u.c. and 5 u.c. samples, is attributed to the STO substrate.

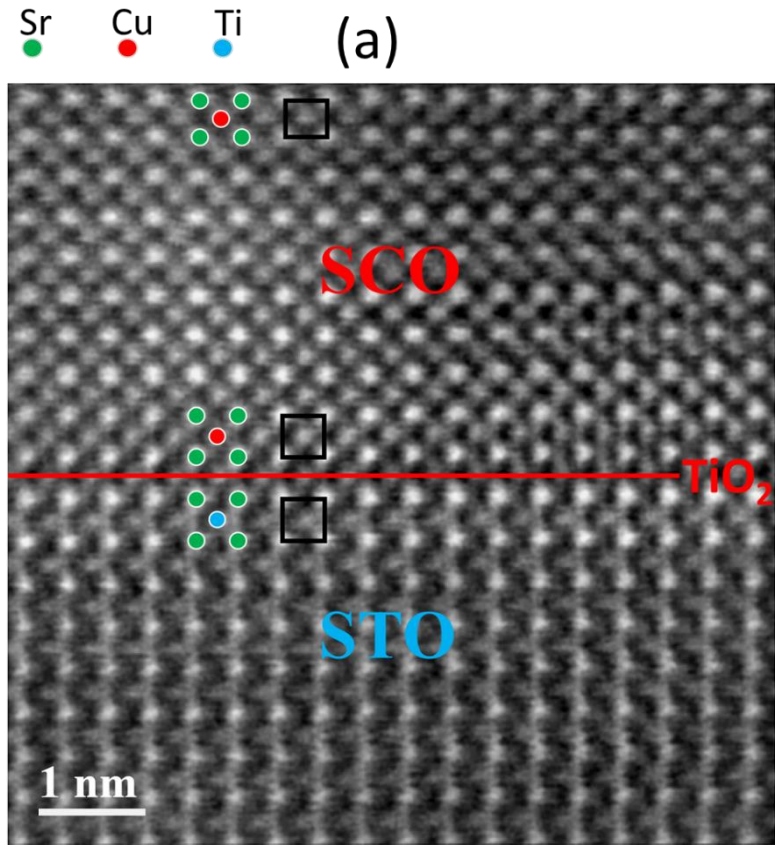

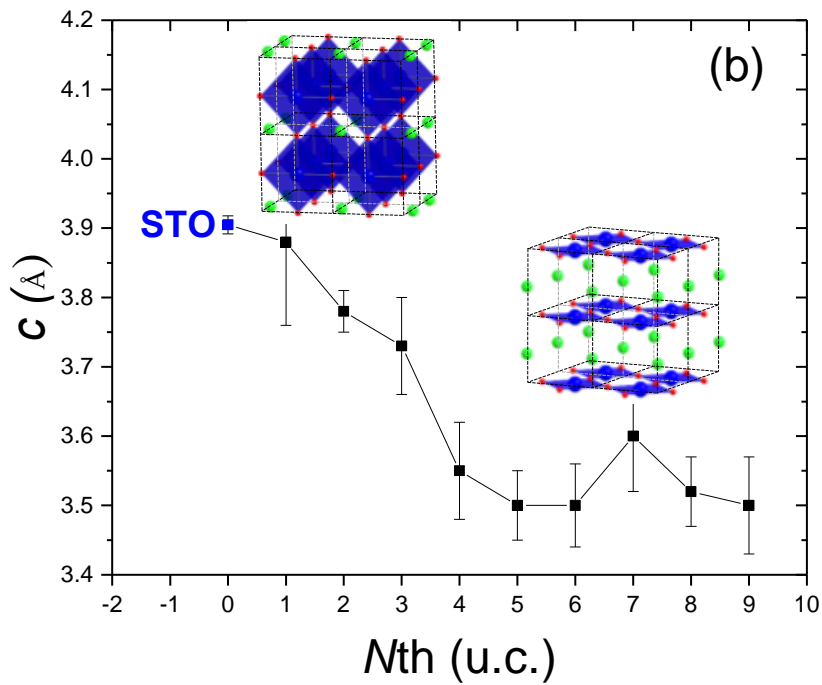

Fig. S7 **STEM image and the measured *c*-axis lattice constant.** The scanning transmission electron microscopy (STEM) images were taken using JEOL ARM200F equipped with a spherical aberration (Cs) corrector at 200 kV accelerating voltage.

The semi convergent angle was 25 mrad, which formed a < 1 Å electron probe. The semi collection angle of high-angle annular dark-field (HAADF) detector was 68 to 280 mrad. (a) Cross section view of HAADF image of SCO/STO. The $TiO_2$-terminated interface is marked by the horizontal red line. The STEM image clearly shows that SCO close to the STO interface forms a chain type structure similar to STO. The similarity is highlighted by the *square* black boxes of the *same* size at the center of this cross-section view. On the other hand, SCO away from the interface forms the planar type structure with a smaller *c*-axis than those of SCO and STO. This is marked by a *rectangular* black box on the top part of image. (b) The *c*-axis spacing between SCO layers. The first layer is closest to the interface. It shows $c = 0.388$ nm, which is slightly smaller than $c = 0.3905$ nm of the STO substrate. The *c*-axis spacing suddenly drops to $c \approx 0.350$ nm after the first three layers. It is noted that this value is already very close to the *c* value of 0.343 nm in bulk planar type SCO. We thus denote that the first three SCO layers have the chain type structure whereas the SCO layers that are far ways from STO interface (roughly 5 u.c.) have the planar structure. Overall, the STEM and Laue nano-diffraction (Fig. S7 and Fig. 4(b)) results support the model depicted in Fig. 4 (c).

**Note 1. The number of conducting $CuO_2$ layers**

The number of $CuO_2$ layers participating in the electrical conductivity in the present case remains a puzzle. A hint about this issue comes from the spectral weight of ZR singlets in Fig. 4(b). If the total contribution arises from a single $CuO_2$ plane, such a significant spectral weight would unlikely be observed. The metallic layers are certainly covered on top by the planar-type SCO that is a Mott insulator, as depicted in Fig. 6. This point is based on the following facts. The peak of the UHB in the O *K*-edge XAS is scarcely observable for $p = 0.22$ and beyond, due to the spectral weight transfer (29,30). If all planar SCO layers were over-doped and metallic, one would not have observed such significant UHB spectral weight for the metallic SCO/STO shown in Fig.4.